\documentclass[epjCONF]{svjour}
\usepackage{graphics}
\usepackage[varg]{txfonts} 
\usepackage[latin1]{inputenc}
\usepackage{natbib}
\session-title{Conference Title, to be filled}
\begin{document}

\title{An abundance study of the red giants in the seismology fields of the {\it CoRoT} satellite}
\author{T. Morel\inst{1}
\and A. Miglio\inst{2} 
\and N. Lagarde\inst{3}
\and J. Montalb\'an\inst{1} 
\and M. Rainer\inst{4} 
\and E. Poretti\inst{4} 
\and S. Hekker\inst{5} 
\and T. Kallinger\inst{6} 
\and B. Mosser\inst{7} 
\and M. Valentini\inst{1} 
\and F. Carrier\inst{6} 
\and M. Hareter\inst{8} 
\and L. Mantegazza\inst{4}
\and J. De Ridder\inst{6}}
\institute{
Institut d'Astrophysique et de G\'eophysique, Universit\'e de Li\`ege, Belgium
\and 
School of Physics and Astronomy, University of Birmingham, UK
\and
Geneva Observatory, University of Geneva, Versoix, Switzerland
\and 
INAF -- Osservatorio Astronomico di Brera, Merate, Italy
\and 
Astronomical Institute 'Anton Pannekoek', University of Amsterdam, The Netherlands
\and 
Katholieke Universiteit Leuven, Instituut voor Sterrenkunde, Leuven, Belgium
\and 
LESIA, CNRS, Observatoire de Paris, Meudon, France
\and 
Institute for Astrophysics, University of Vienna, Austria
}
\abstract{
A precise characterisation of the red giants in the seismology fields of the {\it CoRoT} satellite is a prerequisite for further in-depth seismic modelling. The optical spectra obtained for 19 targets have been used to accurately estimate their fundamental parameters and chemical composition. The extent of internal mixing is also investigated through the abundances of Li, CNO and Na (as well as $^{12}$C/$^{13}$C in a few cases). 
} 
\maketitle
\section{Introduction} \label{sect_intro}
Observations of red giant stars by space-borne observatories such as {\it Kepler} or {\it CoRoT} offer for the first time the opportunity to derive some fundamental properties of these stars from the modelling of their solar-like oscillations [\citealp{christensen_dalsgaard11}]. Accurate estimates of both the seismic observables and the non-seismic constraints (e.g., effective temperature, chemical composition) not only pave the way for a successful modelling of the space data but also allow one to better interpret the abundance results thanks to the knowledge of the mass and evolutionary status, for instance. We present here the first salient results of an abundance study of the red giants lying in the {\it CoRoT} seismology fields. A similar study concentrating on stars in the exoplanet fields is presented by Valentini et al. (these proceedings).

\section{Targets, observations and methods of analysis}\label{sect_targets}
Our sample is made up of 19 red giants, among which 14 have already been quasi-continuously observed by the {\it CoRoT} satellite during runs lasting between about 50 and 150 days. The observations of HD \nolinebreak 50890 and HD \nolinebreak 181907 are discussed by [\citealp{baudin12}] and [\citealp{carrier10}], respectively. Three stars are likely members of the open cluster NGC 6633 based on their radial velocities.
Five bright, well-studied red giants were also observed to validate the analysis procedures. High-resolution optical spectra of the {\it CoRoT} targets (either HARPS or FEROS) were acquired as part of the ground-based follow-up campaigns [\citealp{poretti12}].

The atmospheric parameters ($T_{\rm eff}$, $\log g$, microturbulence) and abundances of 12 metals (Fe, Na, Mg, Al, Si, Ca, Sc, Ti, Cr, Co, Ni and Ba) were self-consistently determined from the spectra using a classical curve of growth analysis. On the other hand, the abundances of Li, C, N and O (as well as the $^{12}$C/$^{13}$C isotopic ratio for 4 stars) were derived from spectral synthesis of atomic or molecular features. In each case, Kurucz atmosphere models and the line-analysis software MOOG were used. Excitation and ionisation equilibrium of iron were used to derive $T_{\rm eff}$ and $\log g$, while the microturbulence was inferred by requiring no dependence between the Fe I abundances and the line strength. A comparison with literature data, temperatures estimated from interferometric data for the five benchmark stars or surface gravities derived from seismic scaling relations supports the reliability of our results. The analysis has been repeated after fixing the gravity to the likely more accurate seismic value where possible. Although the results always agree within the error bars, this is expected to improve both the levels of accuracy and precision. Because not all stars have a seismic gravity available, we only discuss below the results based on the spectroscopic values. However, our conclusions remain unchanged.

\section{Preliminary results}
\begin{figure}[ht]
\begin{minipage}[t]{0.46\linewidth}
\centering
\resizebox{1.0\columnwidth}{!}{%
\includegraphics{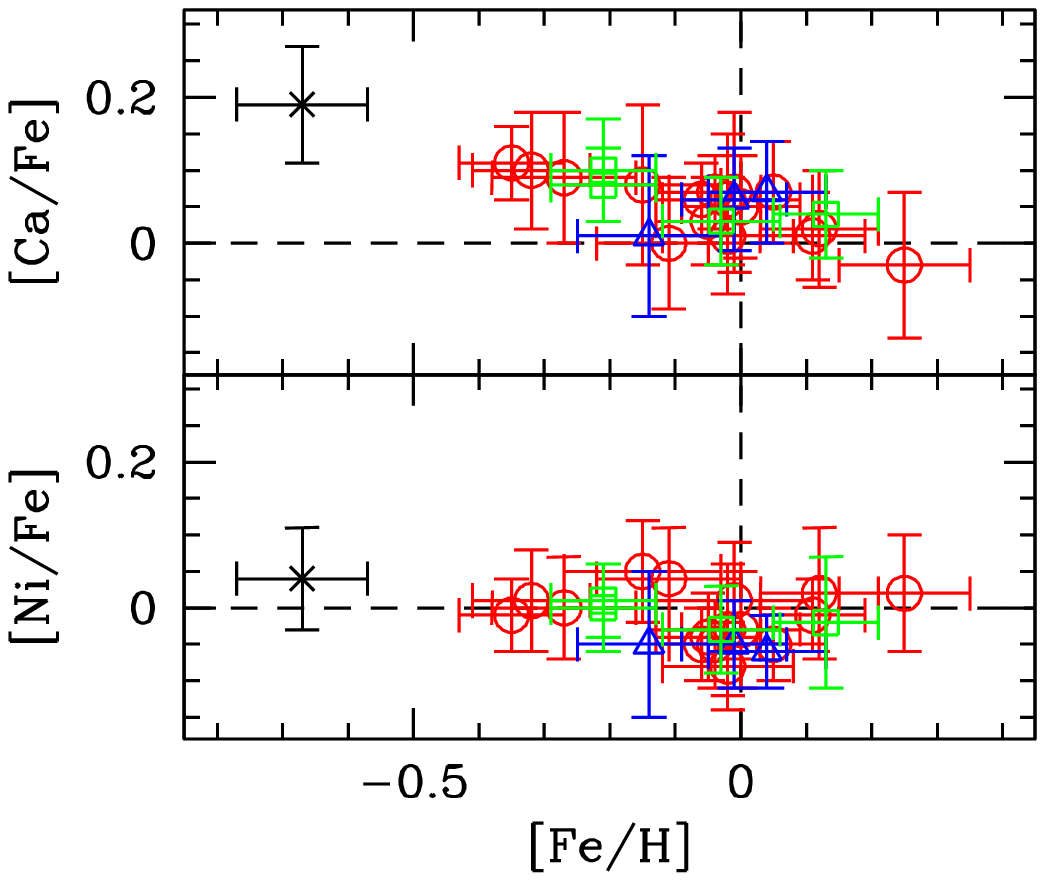}}
\caption{Abundance ratios of Ca and Ni with respect to iron, as a function of [Fe/H] (circles: {\it CoRoT} targets, triangles: stars in NGC 6633, cross: $\alpha$ Boo, squares: the four other stars used for validation).}
\label{fig_patterns}
\end{minipage}
\hspace{0.5cm}
\begin{minipage}[t]{0.46\linewidth}
\centering
\resizebox{1.0\columnwidth}{!}{%
\includegraphics{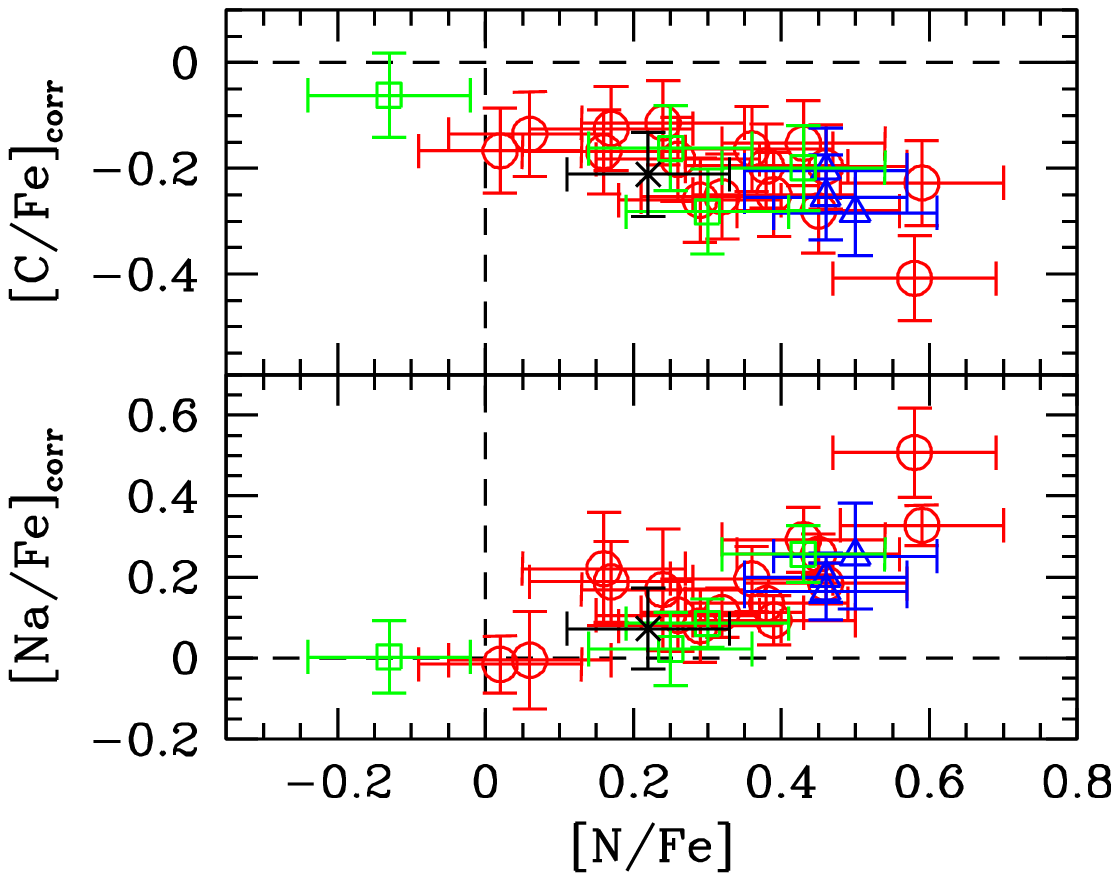}}
\caption{Abundance ratios of C and Na with respect to iron after correction for the chemical evolution of the Galaxy, as a function of [N/Fe] (same symbols as in Fig.\ref{fig_patterns}).}
\label{fig_mixing}
\end{minipage}
\end{figure}

The behaviour of several elements mainly reflects the chemical evolution of the Galaxy. For instance, the abundance of the $\alpha$-element Ca relative to Fe increases as the metallicity decreases, whereas the iron-peak element Ni closely follows Fe (Fig.\ref{fig_patterns}). On the other hand, the variations of the abundances of some elements (e.g., CNO) are not only the result of the nucleosynthesis history of the ISM, but also of internal mixing phenomena. To obtain abundances that are to first order free of the effects related to the former, we have removed the trend as a function of [Fe/H] found in dwarfs of the Galactic thin disk for C, O, Na and Al [\citealp{ecuvillon04b}, \citealp{ecuvillon06}, \citealp{reddy03}] (values appropriate for thick-disk stars were used for the standard star $\alpha$ Boo). No corrections were applied to the N abundances, as no such trend is discernible [e.g., \citealp{reddy03}]. The well-defined correlations observed between these corrected abundances (Fig.\ref{fig_mixing}) will be used to study the mixing processes operating within our sample (e.g., thermohaline instabilities and rotation). 

\begin{thebibliography}{}
\bibitem{christensen_dalsgaard11} Christensen-Dalsgaard J., to be published in \textit{Asteroseismology}, Canary Islands Winter School of Astrophysics, Volume XXII, (editor P. L. Pall\'e, Cambridge University Press), arXiv:1106.5946
\bibitem{baudin12} Baudin F. et al., A\&A \textbf{538}, (2012) A73
\bibitem{carrier10} Carrier F. et al., A\&A \textbf{509}, (2010) A73
\bibitem{poretti12} Poretti E. et al., to be published in Proceedings of the 20th Stellar Pulsation Conference Series: 'Impact of new instrumentation and new insights in stellar pulsations', arXiv:1202.3542
\bibitem{ecuvillon04b} Ecuvillon A. et al., A\&A \textbf{426}, (2004) 619 
\bibitem{ecuvillon06} Ecuvillon A. et al., A\&A \textbf{445}, (2006) 633 
\bibitem{reddy03} Reddy B. E. et al., MNRAS \textbf{340}, (2003) 304 
\end{thebibliography}
\end{document}